\documentclass[aps,prb,twocolumn,superscriptaddress,10pt,floatfix]{revtex4-2}
\usepackage{amsmath,amsfonts,amssymb,bm}
\usepackage{graphicx} 
\usepackage{multirow,array}
\usepackage{braket}
\usepackage{hyperref,epsfig,wrapfig}
\hypersetup{colorlinks=true,    linkcolor=blue,    citecolor=blue,     urlcolor=blue}

\begin{document}
\title{
Quantum-to-classical correspondence for spins at finite \\[1mm] temperatures: 
theory and applications
}

\author{A. El Mendili}
\affiliation{Universit\'e Grenoble Alpes, CEA, IRIG, PHELIQS, 38000 Grenoble, France}
\author{M. E. Zhitomirsky}
\affiliation{Universit\'e Grenoble Alpes, CEA, IRIG, PHELIQS, 38000 Grenoble, France}
\affiliation{Institut Laue Langevin, 38042 Grenoble Cedex 9, France}
\date{September 9, 2026}

\begin{abstract} 
We derive a rigorous quantum-to-classical mapping for interacting spin systems at finite temperatures.
Specifically, the large-$S$ asymptotic form of the partition function is given by the partition function
of a classical model for vectors of length $S_C=\sqrt{S(S+1)}$. Quantum corrections to the asymptotic 
result form a series  in powers of $1/[S(S+1)]$.  We calculate the leading temperature-independent 
quantum correction to the magnetic anisotropy constant. The established mapping justifies the use of classical Monte 
Carlo simulations for realistic magnetic Hamiltonians, including anisotropic and frustrated interactions.
As an application, we compute the Curie and N\'eel temperatures for a range of magnetic materials with 
known microscopic interaction constants.
The obtained transition temperatures are in good agreement with measured values. 
Monte Carlo results for the magnetic susceptibility of the collinear 
antiferromagnet MnF$_2$ above and below the transition are also compared with experimental data.
\end{abstract}
\maketitle

\section{Introduction}
\label{Intro}

The ever-growing demand for new magnetic technologies fuels  interest in the theoretical modeling of 
functional magnetic materials across a broad range of temperatures. In recent years, significant progress 
has been achieved in the experimental determination of microscopic interactions in magnetic solids 
through spectroscopic measurements
\cite{Coldea2002,Zhao2008,Savary2012,Jacobsen2018,Wildes2020,Bai2021,Calder2021,Chen2021,Scheie2022,%
Gu2022,Cho2023,Wildes2023,Liu2024,Morano2024,Maksimov2025}
and in their theoretical prediction using first-principal methods \cite{Riedl2019,Szilva2023}.
The thermodynamics of simple quantum magnets  can be studied using powerful quantum Monte Carlo (MC) 
methods \cite{Sandvik2010,Gubernatis2016}. 
Nevertheless, many realistic spin models exhibit some form of magnetic frustration,  which stems from either 
the lattice geometry or competing interactions. Magnetic frustration gives rise to the negative sign problem that 
precludes efficient quantum simulations \cite{Gubernatis2016}. Conversely, classical MC simulations 
\cite{Landau2000} can easily incorporate the full complexity of microscopic interactions. 
This situation raises anew the question of  the validity of the classical approximation for quantum 
spin models.

From a general standpoint \cite{Brussaard1957,Yaffe1982}, a quantum spin system is expected to behave more 
classically as the spin quantum number $S$ increases. The Heisenberg  model 
\begin{equation}
\hat{\cal H} = \sum_{\langle ij\rangle} J_{ij}\, \bm{S}_i\cdot\bm{S}_j - H \sum_i S_i^z 
\label{Hex}
\end{equation}
is a commonly studied spin model in magnetism. Here $\bm{S}_j$ are spin-$S$ operators, $J_{ij}$ are coupling 
constants, ${\langle ij\rangle}$ denotes a sum over interacting spin pairs, with each pair counted once,  
and $H$ is a dimensionless magnetic field.

 A standard approach to reaching the classical limit  in  (\ref{Hex}) is to define
the new operators $\bm{s}_j = \bm{S}_j/S$ \cite{Fisher1964,Millard1971,Conlon1990}, which satisfy 
\begin{equation}
\bigl[ s^\alpha_j, s^\beta_k\bigr] =  \frac{i}{S}\, \varepsilon^{\alpha\beta\gamma}s_j^\gamma \delta_{jk} \ .
\label{sasb}
\end{equation}
In the limit $S\to\infty$, the commutators vanish and the rescaled spins  $\bm{s}_j$ behave as 
classical unit vectors.
Rewriting Eq.~(\ref{Hex}) in terms of $\bm{s}_j$ yields
\begin{equation}
\hat{\cal H} = S^2\sum_{\langle ij\rangle} J_{ij}\, \bm{s}_i\cdot\bm{s}_j - HS \sum_i s_i^z \,.
\label{Hexc}
\end{equation}
Accordingly, the classical model parameters  are
\begin{equation}
J_C =  J S^2 \,,\quad  H_C  =  HS \,.
\label{JHcl}
\end{equation}

Spin-wave theory provides an alternative prospective on the quantum-to-classical mapping at zero temperature.
In an ordered magnetic state, the starting point of the bosonic spin-wave approach  \cite{Holstein1940,Kubo1952,Auerbach}
 is to represent quantum spins by classical vectors of length $S$.
This correspondence is manifested in two exact results: the ground-state energy of the Heisenberg ferromagnet is proportional 
 to $-JS^2$, while the  transition field to the fully polarized state for the Heisenberg antiferromagnet at $T=0$ is proportional to $JS$, 
 both in agreement with the classical scaling (\ref{JHcl}).
 The path-integral formulation of spin-wave theory \cite{Auerbach} gives essentially the same $S$ dependence 
 for basic physical properties.

The problem of the quantum--to-classical correspondence for statistical properties of spin systems
was first addressed by Elliott Lieb in 1973 \cite{Lieb1973}.  He obtained upper and lower bounds on 
the normalized partition function $Z_Q(S)$.  For a linear dependence of the Hamiltonian $\hat{\cal H}$ 
on individual spins, as, e.g., in Eq.~(\ref{Hex}), the bounds are expressed via the partition function 
$Z_C(x)$ of the same model for classical spins of length $x$:
\begin{equation}
Z_C (S) \leq Z_Q(S)  \leq Z_C(S+1)\ .
\label{LiebZ}
\end{equation}
These inequalities, in turn, provide the free-energy bounds 
\begin{equation}
F_C (S+1) \leq F_Q(S)  \leq F_C(S) \,.
\label{LiebF}
\end{equation}
A simple corollary of Eq.~(\ref{LiebF}) is that the classical free energy $F_C(x) $ of the Heisenberg model (\ref{Hex}) 
is a monotonically decreasing function of spin length $x$. Therefore, the optimal length $S_C$ of  classical spins 
representing a quantum system at fixed temperature must satisfy 
\begin{equation}
S\leq S_C \leq S+1 \ .
\label{Sc}
\end{equation}
At $T=0$, spin-wave theory gives $S_C =S$, corresponding to the lower bound.  However, this is not necessarily
 the best choice for arbitrary $T$. 
For example,  in order to match the Curie tail in magnetic susceptibility, the classical spins have to be chosen with 
$S_C = \sqrt{S(S+1)}$, which also agrees with the condition (\ref{Sc}).

The empirical  representation of quantum spins with classical vectors of length $S_C =\sqrt{S(S+1)}$ was
tested in several numerical  studies  \cite{Elstner1995,Oitmaa2004,Engelhardt2006,Johnston2011,Oitmaa2018,Wang2020}.
 In particular, Oitmaa and Zheng \cite{Oitmaa2004} 
calculated transition temperatures of Heisenberg ferromagnets and antiferromagnets on simple- and body-centered 
cubic lattices from the  high-temperature series 
 for $S=1/2$, 1,  3/2. Their results can be fitted to
\begin{equation}
T_c(S) = c_0 S(S+1) + c_1 + \frac{c_2}{S(S+1)} + \ldots 
\label{Tc}
\end{equation}
with model-dependent coefficients $c_n$. 
The leading-term coefficient $c_0$ coincides with the transition temperature of the corresponding classical model with unit-length spins, 
while the remaining terms represent quantum corrections. These corrections appear to be small, less than 3--5\%\ for
$S=3/2$.

The quantum-classical correspondence also applies to low-dimensional systems without finite-temperature phase transitions.
The quantum MC simulations  of the Heisenberg antiferromagnet on a square lattice  
showed that the quantum and classical susceptibilities  $\chi(T)$  and specific heats $C(T)$  agree well for
 $S\geq 3/2$ at  $T\agt 0.8JS(S+1)$, 
 once  the substitution $S_C=\sqrt{S(S+1)}$ is used for classical spins \cite{Johnston2011}. 
 The similarity between static quantum and classical spin-spin correlations, in both magnitude and spatial dependence,
 was also demonstrated in numerical studies
 of various two- and three-dimensional spin models
 \cite{Elstner1995,Cuccoli1997,Kulagin2013,Wang2020,Schneider2025}. 

Taken together, the classical modeling of Heisenberg spin systems at finite temperatures, based on the empirical replacement
 $S_C = \sqrt{S(S+1)}$, is now widely accepted in the literature. Still, in the absence of rigorous results, 
the alternative substitution  $S_C=S$ is also sometimes used, see, e.g.,  
\cite{Cepas2005,Barker2019,Torelli2019,Kartsev2020,Olsen2021}.
According to Eq.~(\ref{Tc}), $T_c$ in this case is underestimated by the factor $S/(S+1)$, 
which amounts to a 22\%\ difference, even for  the relatively large $S=7/2$.
Furthermore, prior numerical studies of the quantum-to-classical
correspondence have focused on Heisenberg spin models and have not considered
the mapping for  anisotropic interactions and the Zeeman energy, although
these terms are essential in most comparisons between theory and experiment.

In our work we revisit the problem of quantum-classical mapping for lattice spin models with arbitrary interactions.
We derive the asymptotic form of the partition function $Z_Q(S)$ for large $S$, Eq.~(\ref{ZQC}), which 
establishes a rigorous basis for representing quantum spins at finite temperatures with classical vectors 
of length $S_C = \sqrt{S(S+1)}$. 
This result provides a theoretical basis for the leading $S(S+1)$ scaling of $T_c$ in
 the empirical formula  (\ref{Tc}) beyond the isotropic nearest-neighbor spin models studied numerically so far \cite{Oitmaa2004,Oitmaa2018}.
We  demonstrate the capabilities of the classical approach by calculating transition temperatures 
for several topical magnetic materials with known microscopic interactions and 
moderate spin values  $S=3/2$, 2, 5/2, and 7/2.
The  transition temperatures obtained from the classical MC simulations agree well with the experimental values.
Thus, comparison of  computed $T_c$ with the   experimental value provides
as a stringent test for any microscopic set of interaction parameters for the spin Hamiltonian of a 
magnetic material.

The paper is organized as follows. 
Section~\ref{QCM} presents our central result:  quantum-to-classical mapping 
for a system of interacting  quantum spins. Section~\ref{LargeS} 
describes the analytic derivation of the asymptotic large-$S$  form 
of the partition function. 
In Sec.~\ref{QCorr} we compute a
 temperature-independent quantum correction to the magnetic anisotropy constant
relevant in the paramagnetic phase above $T_c$.
Section~\ref{MCSim} sums up the classical Monte Carlo simulations performed in this study. 
The effective classical model is elaborated on in Sec.~\ref{EffectiveCM}, Sec.~\ref{MCA} outlines
the MC algorithm, and Sec.~\ref{TT} contains results on the transition temperatures of various materials 
simulated in this study as well as the magnetic susceptibility results for MnF$_2$.
Conclusions are presented in Sec.~\ref{Sum}. Additional information is included 
in appendices: Appendix~\ref{Trace} gives an alternative proof of the quantum-to-classical correspondence 
for spin traces, Appendix~\ref{Marsaglia}  presents a generalization of Marsaglia algorithm used in 
the MC simulations, and Appendix~\ref{Models} describes in detail  spin models for each of the studied 
materials.

\section{Quantum-to-classical mapping} 
\label{QCM}
\subsection{Large-\boldmath{$S$} limit of the partition function}  
\label{LargeS}

Consider a system of $N$ spins with quantum number $S$ described by a Hamiltonian $\hat{\cal H}$. It is  convenient 
to define the normalized partition function 
\begin{equation}
Z =  \frac{1}{Z_0}  {\rm Tr}\,\bigl\{ e^{-\beta \hat{\cal H}}\bigr\}\,,\quad Z_0 =  (2S+1)^N  \ ,
\label{Z}
\end{equation} 
where $\beta= 1/T$. 
As $T\to\infty$, the partition functions of the quantum and classical models
approach the same value, enabling a direct comparison at finite temperatures.

The standard approach to computing  the partition function $Z$ is the high-temperature series expansion  
in powers of $\beta$ 
\begin{equation}
Z = \sum_{n=0}^\infty \frac{(-\beta)^n}{n!}  
\bigl\langle \hat{\cal H}^n\bigr\rangle_0\,,
\label{Zexp}
\end{equation} 
where $\langle\ldots\rangle_0 = (1/Z_0) {\rm Tr}\,\{\ldots\}$  denotes averaging over the uncorrelated 
paramagnetic state at $\beta =0$ \cite{Rushbrooke1974,Oitmaa2006,Tang2013}. Below, we first  recapitulate the basic
properties of the high-temperature series (\ref{Zexp}) and then derive the asymptotic form of
$Z_Q(S)$.

The moments of $\hat{\cal H}$ in Eq.~(\ref{Zexp}) are expressed as sums over finite, connected or disconnected lattice graphs.
These graphs encode different magnetic interactions included in the spin Hamiltonian.
The contribution of each graph is split into two factors: a combinatorial lattice-embedding constant 
and an appropriate multi-spin correlator.  Evaluated 
at $\beta =0$, the correlators further factorize into traces of
products of spin operators with the same lattice index:
\begin{equation}
K_n(S)  = \frac{1}{2S+1} {\rm Tr}\bigl\{  S_i^{\alpha_1} S_i^{\alpha_2}\ldots S_i^{\alpha_n}\! \bigr\} \,.
\label{Kn}
\end{equation}
Number and  sequence of operators in each trace are determined by a lattice graph. At the same time,
the distinction between quantum and equivalent classical models stems solely from the trace values.  

To determine the  large-$S$ limit of the  partition function $Z_Q(S)$, we must explore
the asymptotic behavior of the individual traces $K_n(S)$.  Let us begin with the simplest type of
traces containing only one spin component
\begin{equation}
I_n(S)  = \frac{1}{2S+1} {\rm Tr}\bigl\{ (S^z)^n \bigr\} =  \frac{1}{2S+1}\!\sum_{m=-S}^{S}m^n  \  ,
\label{In}
\end{equation}
which are nonzero for even  $n$. The  sums of powers for integer or half-integer $m$ can be computed 
using Bernoulli polynomials $B_n(x)$, see, e.g.,  \cite{Oldham2009}.
These are defined by the generating function:
\begin{equation}
\frac{te^{xt}}{e^t - 1}  = \sum_{n=0}^\infty  B_{n}(x)\, \frac{t^n}{n!}  \ .
\label{Bn}
\end{equation} 
It follows from (\ref{Bn}) that $B_{n}(x)$ is a polynomial of degree $n$ with a leading term coefficient of 1.  
Bernoulli polynomials obey
\begin{equation}
B_{n}(x+1)-B_{n}(x) = nx^{n-1}\,.
\label{BnD}
\end{equation}
This identity allows us to rewrite (\ref{In}) as a telescoping sum,  with the end result
\begin{equation}
I_n(S)  =  \frac{B_{n+1}(S+1) - B_{n+1}(-S)}{(n+1)(2S+1)}   \,.
\label{IBn}
\end{equation}
Since the numerator is an odd function of  $Y=S+1/2$,  $I_n(S)$ must be an even polynomial in $Y$. 
By substituting $Y^2 = X+ 1/4$, we obtain
\begin{equation}
I_n(S) = \frac{X^{n/2}}{n+1}  + \sum_{k=1}^{n/2-1} a_k X^{k} \,, \quad X = S(S+1) \,,
\label{InSQ}
\end{equation} 
with $a_0=0$, because of  
$B_{n}(1) = B_{n}(0)$ for $n>1$, see (\ref{BnD}).
The asymptotic form of  $I_n(S)$ in the $S\to\infty$ limit is determined  by the 
first term with the highest power of $X$. 
 Another proof of (\ref{InSQ}) based only on elementary functions
is given in Appendix~\ref{Trace}. 

The classical average corresponding to the discrete sum (\ref{In}) is represented by a spherical
 integral 
\begin{equation}
I_n^C(S_C) =  \int\frac{\sin\theta \,{\rm d}\theta\,{\rm d}\varphi}{4\pi} \,(S_C\cos\theta)^n = \frac{S_C^{\,n_{}}}{n+1} \,.
\label{InSC} 
\end{equation}
Notably, the coefficient in (\ref{InSC}) is precisely the same as for the highest-power term in the polynomial
representation of $I_n(S)$.
Hence, for  classical spins of length $S_C = \sqrt{S(S+1)}$,  the average  $I_n^C$ will coincide with
the large-$S$ limit of the quantum trace  (\ref{InSQ}). 

We must now  prove  that the same classical representation applies to the  asymptotic form of an arbitrary trace 
$K_n(S)$. Note that extra terms arising from the commutation of spin operators have a reduced number of $S$ 
factors and can be disregarded in the large-$S$ limit. Next,  $S^{x}$ and $S^{y}$ are substituted with 
$S^\pm = S^x\pm iS^y$. Rotation symmetry about the $z$-axis requires an equal number of $S^+$ and 
$S^-$ in every nonzero trace.  Furthermore, moving $S^+$ and $S^-$ along the trace and replacing the adjacent 
operators with $S^+ S^- \approx S(S+1) - (S^z)^2$ one can recursively eliminate transverse spin components. 
Then,  the highest power term for $K_n (S)$ can be represented by a sum over $I_p (S)$ with $p\leq n$ multiplied 
by $X^{(n-p)/2}$. A somewhat tedious but straightforward generalization of the  above arguments shows that 
a polynomial representation similar to Eq.~ (\ref{InSQ}) holds for any trace $K_n (S)$ \cite{Bose1964}.

Now, consider a spherical integral representing $K_n (S)$ for classical spins. Using the same steps as before 
and $S^+ S^- = S_C^2 - (S^z)^2$, one can reduce the integral to a linear combination of classical integrals 
$I_p^C(S_C)$ with $p\leq n$ multiplied by powers of $S_C$. Due to the correspondence between (\ref{InSC}) 
and the leading term in (\ref{InSQ}), we can now conclude that the large-$S$ limit of any trace of products of spin 
operators (\ref{Kn}) is given by its spherical-integral  representation  with classical vectors of length $S_C = \sqrt{S(S+1)}$. 
In addition, corrections to the asymptotic behavior form a series in powers of $1/S(S+1)$.

After proving the quantum-classical correspondence for  traces of products of spins, we can now establish a similar 
result for the partition function. Consider a quantum spin model and its classical equivalent with $S_C = \sqrt{S(S+1)}$. 
The two models have identical lattice graph representations
for each term in the $\beta$ expansion (\ref{Zexp}). Because of the correspondence
between quantum and classical correlators, the $n$th order term for the classical model in (\ref{Zexp})
is equal to the large-$S$ contribution of the $n$th order quantum term.
Then, the series resummation yields
\begin{equation}
Z_Q(S) = Z_C(S_C) \Bigl[ 1+ O\Bigl( \frac{1}{S_C^2} \Bigr)\Bigr],  \quad  S_C^{\,2_{}}=S(S+1)\,,
\label{ZQC}
\end{equation}
where the second term  on the rhs represents a quantum correction.   
As long as Eq.~(\ref{ZQC}) holds, the free energy and its derivatives allow for similar expansions 
in powers of $1/S_C^2$ with the classical result corresponding to the leading term.

Although, the above derivation is based on the $\beta$
 expansion, the asymptotic form of the partition function
 is essentially a non-perturbative result, since the quantum-classical correspondence holds to all orders in $\beta$.
Still, convergence of the whole series is necessary to perform the resummation leading to  (\ref{ZQC}).
The domain of convergence for $Z$ as a function of $\beta$
usually extends from $\beta=0$ 
to the nearest physical critical point $\beta_c$ ($T_c$).
Consequently, the asymptotic form  
is exact for  $T>T_c$.  However, a finite radius of convergence  does not imply
an immediate breakdown of the classical description below $T_c$.
Renormalization group theory predicts that  both critical exponents and universal amplitude ratios  
for physical observables are fully determined by the underlying universality
class  of the critical point \cite{Pelissetto2002}. Hence, if the equivalent classical model provides a
quantitative description of the quantum model for  $T>T_c$,
the same correspondence must also hold in the critical region at $T<T_c$. 

We conclude this subsection by noting  that representation of the spin traces (\ref{Kn}) and (\ref{In}) as polynomials in 
$X=S(S+1)$ is well known in the literature, see, for example, 
\cite{Rushbrooke1974,Ambler1962,Bose1964,Dalton1968,Subramanian1974,Meyer1978,Schmidt2011,Lohmann2014}. 
From a symmetry point of view, this polynomial form follows
from the fact that spin traces are rotationally invariant and that the $SO(3)$ group has a single Casimir operator 
$\hat{\bm{S}}^2 = S(S+1)$ \cite{Zee2016}.
However, to the best of our knowledge,
neither the universal matching between the leading term
of $I_n(S)$ and the classical result (\ref{InSC}) nor the resulting
quantum-classical correspondence for the partition function (\ref{ZQC}) has been
established previously.

\subsection{Quantum Correction to Magnetic Anisotropy}
\label{QCorr}

In the paramagnetic phase above $T_c$, the quantum corrections to the leading asymptotic term
in (\ref{ZQC}) form a series in powers of $1/S(S+1)$ with temperature-dependent 
coefficients. For large spins, this series may converge faster than the standard $1/S$ expansion at $T=0$.
Consequently, quantum spin systems tend to behave more classically at finite temperatures. Consistent with 
this trend, a striking similarity between quantum and classical spin-spin correlators  is observed even
for  $S=1/2$  
\cite{Kulagin2013,Wang2020,Schneider2025}.
Conversely, in the magnetically ordered state below $T_c$, the emergence of long-range order breaks 
the pristine $1/[S(S+1)]$ scaling of quantum corrections, triggering a crossover to the conventional $1/S$ spin-wave expansion at zero temperature. Furthermore, the zero-temperature semiclassical  limit  is not necessarily unique, but may depend  on the presence of 
various multipolar terms in the spin Hamiltonian \cite{Zhang2021,Remund2022,Dahlbom2025}.

A systematic investigation of temperature-dependent quantum corrections both above and below $T_c$ goes
beyond the scope of this work. Nevertheless,
the approach formulated in the preceding subsection allows us to derive a simple quantum correction 
to the anisotropy constant. For a uniaxial system, the lowest-order single-ion anisotropy 
can be written as
\begin{equation}
\hat{\cal H}_a = D \sum_i\Bigl[\bigl(S_i^z\bigr)^2 - \frac{1}{3}\,S(S+1)\Bigr] 
\label{Ha}
\end{equation}
with an energy shift introduced for convenience. 
In the high-temperature series (\ref{Zexp}), each appearance of  $\hat{\cal H}_a$ contributes one factor of $D$.
Hence, terms proportional to $D^p$ involve $p$ insertions of the local anisotropy operator from (\ref{Ha}) and 
an arbitrary number of spin-spin interactions from the other terms in the Hamiltonian. 
Following the procedure outlined in Sec.~\ref{LargeS}, all traces  in the high-temperature expansion 
can be reduced to the computation of
\begin{equation}
I_n^{p}(S)  =\bigl\langle \bigl[(S^z)^2 - {\textstyle \frac{1}{3}}\,X\bigr]^p (S^z)^n \bigr\rangle_0\ ,
\label{Inp}
\end{equation}
where $X=S(S+1)$, and each trace is implicitly associated with a prefactor $D^p$. 

From the discussion in Sec.~\ref{LargeS}, we know that $I_n^{p}(S)$ can be represented as a polynomial in $X$.
The idea is to study its roots for arbitrary $p$ and $n$. Let's start with the simplest case $p=1$.
For $S=1/2$ ($X=3/4$), the expression in the square brackets in Eq.~(\ref{Inp}) is zero. 
As a result, the polynomial $I_n^{1}(S)$) must vanish at $X=3/4$ and, consequently, contains the factor $(X-3/4)$.
Combining this algebraic factor with the single-ion constant $D$ yields the renormalized magnetic anisotropy
\begin{equation}
D_C = D \bigl[S(S+1)-3/4\bigr] \,.
\label{DC}
\end{equation}
The combination (\ref{DC}) is known to appear  in the expression 
for the Curie-Weiss temperature   \cite{Wang1971} as well as in the $\beta^2$ contribution 
to the dynamical structure factor \cite{Malinoski1973}.

The goal now is to demonstrate that the renormalization (\ref{DC}) holds to {\it all orders} in $D$ and $\beta$. Since $\bigl[(S^z)^2 - 1/4\bigr] \equiv 0$ 
is a strict operator identity for $S=1/2$, any joint trace (\ref{Inp}) involving $p$ factors of the local anisotropy operator has 
a root at $X=3/4$ with multiplicity $p$. To prove the correct root multiplicity consider a modified trace $I_n^{p}(S,\varepsilon)$ 
obtained upon a substitution $X \to X + \varepsilon$ into Eq.~(\ref{Inp}) with small $|\varepsilon|\ll 1$.
Its direct evaluation gives $I_n^{p}(1/2,\varepsilon) \propto \varepsilon^p$. 
On the other hand,  the polynomial form of $I_n^{p}(S,\varepsilon)$ implies that
this  small-$\varepsilon$ behavior is possible only if $X=3/4$ is a root of multiplicity $p$:
\begin{equation}
I_n^{p}(S) = (X-3/4)^p f(X) \,.
\end{equation}
Consequently, the factor $(X-3/4)^p$ appears next to $D^p$ throughout the entire high-temperature series, 
thereby proving the renormalization (\ref{DC}) for all $T>T_c$.

The renormalized constant $D_C$ vanishes for $S=1/2$, satisfying the natural physical requirement that 
single-ion anisotropy is absent in spin-1/2 systems.  Straightforward generalization  of the above analysis 
shows that the same quantum correction applies to all second-order
terms in the crystal-field Hamiltonian as, e.g., the bi-axial anisotropy that may appear in low-symmetry
magnetic materials, see Sec.~\ref{CoPS3}.
Including the quantum correction  for $D_C$  in the paramagnetic
phase  is somewhat
similar to the  zero-temperature substitution $D \to D[1-1/(2S)]$ appearing in  spin-wave expression for the anisotropy gap
of quantum ferro- and antiferromagnets  \cite{Oitmaa2008,ElMendili2025}. 
In both cases, a model-independent quantum correction is included initially at the  `harmonic' level, whereas
other corrections of similar order in $1/S$, or $1/S(S+1)$, are computed at the next approximation level. 

Providing the fundamental insight, the mapping to an effective classical model also offers practical benefits. 
The obtained spin system can be simulated using classical MC techniques \cite{Landau2000}. 
 This numerical approach provides a versatile and efficient alternative to explicit calculation 
 of  high-temperature series, which rapidly becomes 
 intractable for realistic spin models with multiple-neighbor exchanges. 
 In the following section, we demonstrate the predictive capability of 
 the quantum-to-classical mapping by presenting MC simulations of realistic spin models 
 for several topical magnetic materials.

\section{Monte Carlo simulations}
\label{MCSim}

\subsection{Effective classical model}
\label{EffectiveCM}

Consider a generic spin-$S$ Hamiltonian with arbitrary exchange interactions, 
the single-ion anisotropy 
and the Zeeman energy:
\begin{equation}
\hat{\mathcal{H}}  =\sum_{\langle ij \rangle}  J_{ij}\, \bm{S}_i\cdot \bm{S}_j +
\sum_i \Bigl[ D{{S}_i^z}^2 - g\mu_B\,\bm{H}\cdot \bm{S}_i \Bigr]  \,.
\label{Hgen}
\end{equation}
Other common interactions, such as the biaxial anisotropy or the biquadratic exchange, 
can also be included.
The classical Monte Carlo (CMC) codes are typically written for unit-length vector spins. Accordingly,
the quantum to classical mapping proceeds as 
\begin{equation}
\bm{S}_i \to \sqrt{S(S+1)}\; \bm{s}_i\,,\quad |\bm{s}_i| = 1\,.
\label{QCMS}
\end{equation}
It gives the following values for the microscopic parameters of the effective classical model:
\begin{eqnarray}
\label{map_MC}
 && J_C  =  JS(S+1)\,, \quad  H_C = g\mu_BH  \sqrt{S(S+1)}\,,     \nonumber \\
 && D_C =  D \bigl[S(S+1) - 3/4  \bigr] \,.
\end{eqnarray}
Based on the arguments in Sec.~\ref{QCorr}, the anisotropy constant  is taken with the quantum correction.
Furthermore, going beyond  Eq.~(\ref{Hgen}), the same substitution (\ref{QCMS}) applies
to every possible term in the spin Hamiltonian, including the anisotropic exchange, the Dzyaloshinskii-Moriya interactions,
 the biquadratic terms, etc. However, the quartic and 
higher-order  single-ion (crystal-field) terms may require reevaluation of the 
subleading quantum corrections, as compared to the results of Sec.~\ref{QCorr}.

In Monte Carlo simulations, it is convenient to use dimensionless spin Hamiltonians.
This is usually achieved by scaling $\hat{\cal H}$ to the largest exchange constant $J_C=JS(S+1)$.
Then, the physical observables are related to  their dimensionless MC values by
\begin{eqnarray}
\label{Phys_MC}
 && k_B T/T_{\rm MC} = JS(S+1)\,, \quad C/C_{\rm MC} = k_B N_0\,,     \\
 && M /M_{\rm MC} = g\mu_B N_0  \sqrt{S(S+1)}\,, \  \ \chi /\chi_{\rm MC} = \frac{(g\mu_B)^2}{J}N_0\,. 
 \nonumber
\end{eqnarray}
Here,  
$C_{\rm MC}$, $M_{\rm MC}$ and $\chi_{\rm MC}$ are the heat capacity, the magnetization and the susceptibility 
normalized per spin, 
$k_B$ is the Boltzmann constant, $\mu_B$ is  the Bohr magneton, and   $N_0$  is  the Avogadro number.

\subsection{Monte Carlo algorithm}
\label{MCA}

The Monte Carlo simulations have been performed on finite  clusters with linear sizes $L$ and periodic 
boundary conditions. The total number of spins in each cluster is $N = N_c\times L^D$, where $N_c$ is 
the number of magnetic sites in the crystal unit cell and $D$ is the dimensionality. The MC results presented 
below were obtained for lattices with up to $N = 5\times10^4$ spins.
The standard Metropolis algorithm has been used for accepting or rejecting  new spin directions. 
To get better statistics, the acceptance rate was kept at the level of 30--50\%\ for all temperatures.
This is usually achieved by restricting variations of $z$-component of spins in the local frame to 
$\Delta S^{\rm max} \simeq T$ \cite{Landau2000}. Restricted spin moves have been implemented using
modified Marsaglia algorithm \cite{Marsaglia1972}, 
see Appendix \ref{Marsaglia} for further details.

One Metropolis sweep over the lattice consists of successive attempts to change orientation of every spin. 
This is followed by microcanonical over-relaxation moves, which further improve  statistics of individual 
MC runs by facilitating a random walk in the phase space \cite{Landau2000}. 
Specifically, the $\pi$ rotations  of spins around their local field directions are used 
starting with a first randomly chosen spin \cite{Kanki2005}. One MC step consists of a Metropolis sweep together 
with five subsequent over-relaxation sweeps. At each temperature, $10^5$ MC steps have been performed 
for equilibration followed by up to $5\times 10^5$ MC steps for measurements.
The individual MC runs have been initialized at a high enough temperature $T_{\rm MC} \simeq 3$--5 by completely
random spin configurations followed by gradual cooling across the simulated temperature range. 
A total of 100--500 independent MC runs were simulated in parallel on a computer cluster. The presented MC results 
along with the statistical errors were obtained by averaging the data from different runs.

During MC runs, we compute averages of the internal energy $E$ and squared amplitudes of the
Fourier harmonics of the spin density 
\begin{equation}
(m_{\bm{Q}}^\alpha)^2 = \frac{1}{N^2} \sum_i \langle s^\alpha_is^\alpha_j\rangle \, 
e^{i\bm{Q}(\bm{r}_i-\bm{r}_j)}
\label{mq}
\end{equation}
for the ordering wave-vector(s) $\bm{Q}$ of a specific magnetic model.
The heat capacity $C$ and the magnetic susceptibility $\chi$ are obtained from  fluctuations of the internal energy and 
the magnetization $M^\alpha$ by
\begin{equation}
C  = \frac{\langle E^2\rangle - \langle E\rangle^2}{T^2}\,,\quad
\chi^{\alpha\alpha}  = \frac{\langle (M^\alpha)^2\rangle - \langle M^\alpha\rangle^2}{T}\,.
\end{equation}
Here, all quantities represent the total values for a simulated cluster.  

\begin{figure}[tb]
\centerline{
\includegraphics[width=0.9\columnwidth]{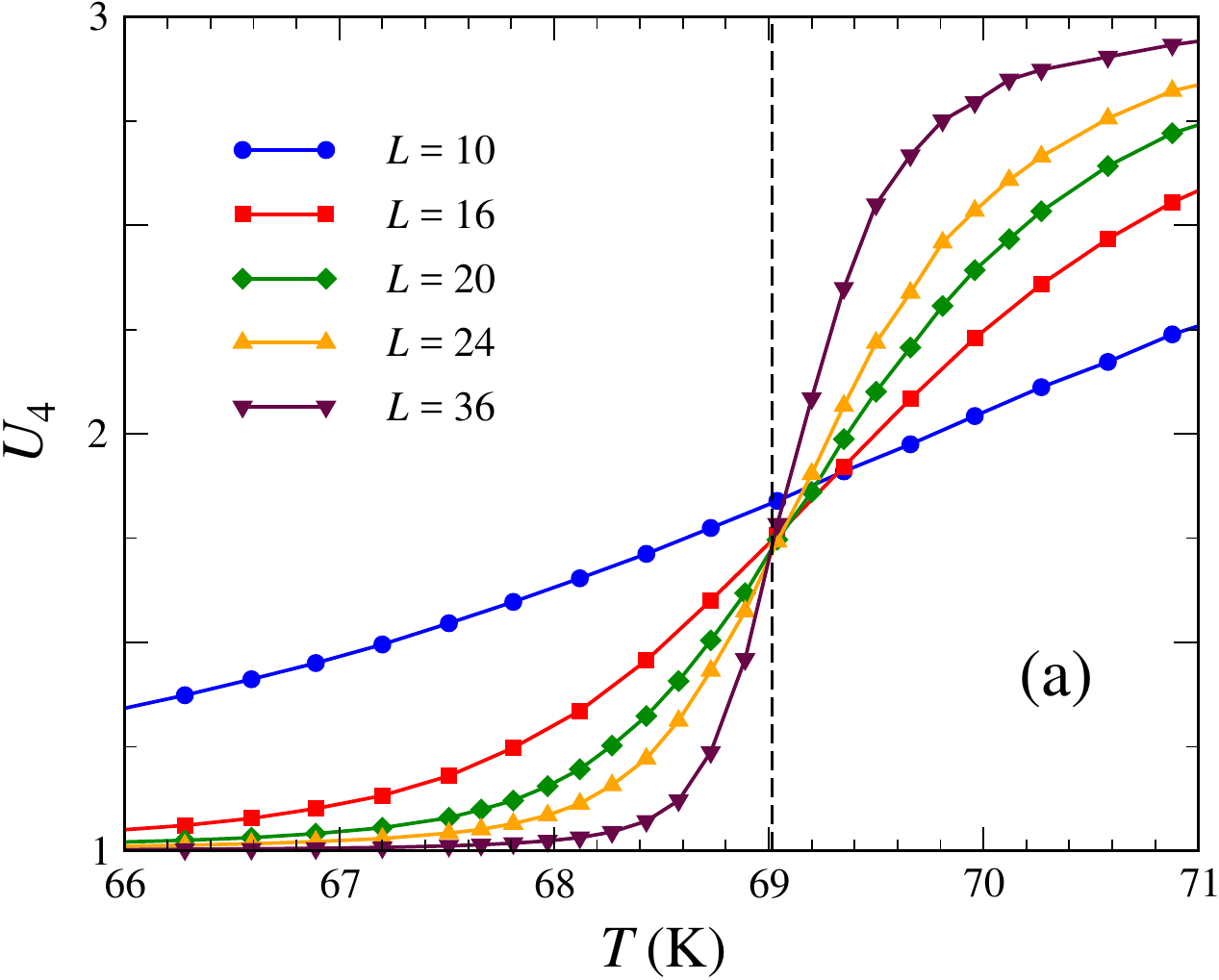}
}
\vskip 7mm
\centerline{
\includegraphics[width=0.9\columnwidth]{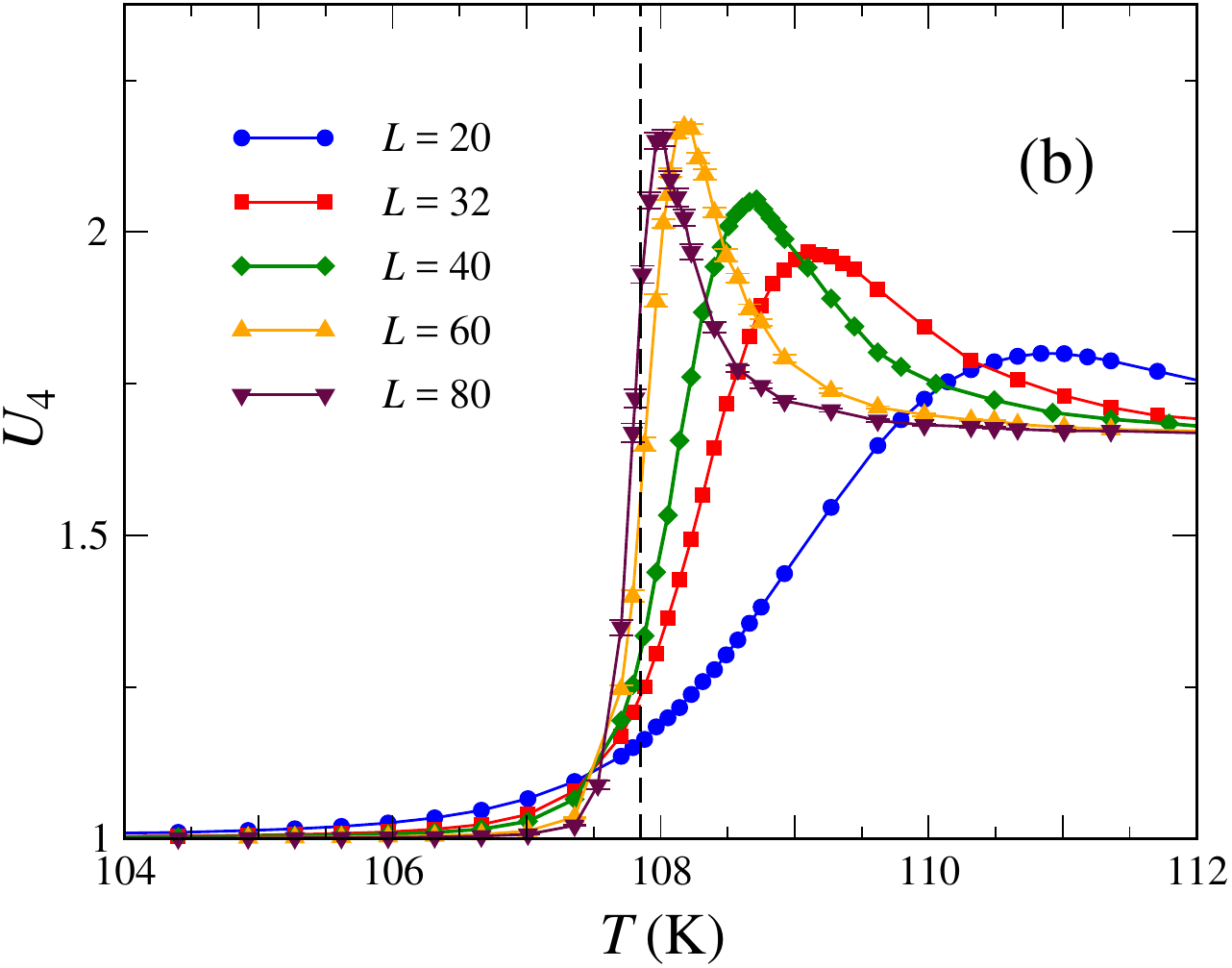}
}
\caption{Fourth-order cumulants versus temperature for the  spin models of  MnF$_2$  (a) and  
FePSe$_3$ (b). Phase transition temperatures  are indicated by the vertical dashed lines.
If not shown, the  error bars are smaller than the symbol sizes.
}
 \label{Cumulants}
\end{figure}

In this subsection we present the results of our Monte Carlo simulations for several magnetic insulators
with  intermediate to large spins, including $S=3/2$, 2, 5/2, and 7/2. Specifically, we studied the following compounds:
CrI$_3$,  CrSBr, CoPS$_3$, 
FePS$_3$, FePSe$_3$, MnPSe$_3$,  Rb$_2$MnF$_4$, MnF$_2$, MnTe,   EuO, and EuS.  
Apart from the current research interest,  these examples were chosen due to detailed information available in 
the literature about their microscopic interaction constants. 
Further details on the simulated spin models, their interaction parameters, along with
the experimental or theoretical methods used to obtain them,  are included in Appendix \ref{Models}.

For each temperature, we also compute the fourth-order cumulant of the order parameter
\begin{equation}
 U_4 =  \frac{ \langle m_{\bm{Q}}^4 \rangle}{\langle m_{\bm{Q}}^2 \rangle^2} \ .
\label{U4}
\end{equation}
In the next subsection, we show how it can be used to locate phase transition points,
see also \cite{Landau2000} for extra details.

\subsection{Monte Carlo results}
\label{TT}

We begin with the transition temperatures.  Due to the rounding effects on finite lattices, the peak position in the specific heat is a rather poor method for locating  $T_c$. The fourth-order cumulants of the order parameter (\ref{U4}) provide instead an accurate approach  that does not require prior knowledge of the critical exponents. The scale-invariance hypothesis suggests the following form of $U_4$ as a function of distance to the critical point $\tau = (T-T_c)/T_c$
 and the system linear size $L$:
 \begin{equation}
 U_4(\tau,L) = f\bigl(\tau L^{1/\nu}\bigr) \ ,
 \label{U4sc}
 \end{equation} 
 where $f(x)$ is a universal function \cite{Landau2000}. 
Therefore, plots of $U_4(T)$  for different system sizes $L$ intersect near $T_c$.
Deviations from the true crossing arise due to small corrections to the above scaling form. 
Below, we give two illustrations for this method.

Figure~\ref{Cumulants}(a) shows the cumulant plot used  to locate the transition temperature 
$T_c$ of MnF$_2$. This collinear easy-axis antiferromagnet on a body-centred tetragonal 
lattice is described by a scalar order parameter $m^z_{\bm{q}}$ with $\bm{q} = (\pi,\pi,\pi)$.  
The spin model of MnF$_2$, see Sec.~\ref{MnF2} for details, have been simulated on
periodic lattices with $N=L^3$ spins and $L = 10$--$36$. For $T\gg T_c$, the cumulants are close 
to $U_4= 3$ consistent with the Gaussian fluctuations of the order parameter. 
Below transition, $m^z_{\bm{q}}$ is finite and 
$U_4  = 1$. The cumulants vary monotonically between these two limiting values 
crossing at  $T_c= 69.01$~K. The crossing point of  $U_4(T,L)$  is rather tight and allows us to
 estimate $T_c$ with high precision without further finite-size scaling of the individual 
crossing points. 

For the second example we choose the van der Waals material FePSe$_3$. This collinear 
honeycomb antiferromagnet has a strong easy-axis anisotropy, which allows us to 
study its phase transition in a simplified 2D spin model, see Sec.~\ref{FePSe3}. 
The in-plane zigzag antiferromagnetic order 
in FePSe$_3$ breaks  the $C_3$ rotation symmetry of the honeycomb lattice and has a three-component
order parameter corresponding to the  $M$ point in the hexagonal Brillouin zone.
Figure~\ref{Cumulants}(b) shows the fourth-order cumulants  obtained by simulating
periodic lattices with $N=2L^2$ spins, $L = 20$--80.
For an $n$-component vector order parameter, above $T_c$ each component 
 fluctuates independently resulting in $U_4 \approx (n+2)/n$. Accordingly,
the cumulants approach 5/3 ($n=3$)  at high temperatures, see  Fig.~\ref{Cumulants}(b).

\begin{center}
\begin{table*}[th]
    \centering
    \begin{tabular}{c|c|c|c|c|c}
\hline  
\qquad {Material}\qquad\qquad 
 &  \quad \qquad  {Spin Model}  \quad  \qquad  \qquad 
 &  \qquad  {Method}   \qquad  \qquad 
 & ~~~$T_c^\text{MF}$ (K)~~~ 
 &  \quad{$T_c^{\text{MC}}$ (K)}\quad\rule{0pt}{2.6ex}  
 & \quad $T_c^{\rm exp}$ (K)\qquad
\\[1mm]
\hline\hline  
EuO, $S=7/2$ 
  & \multirow{2}{*}{$\{J_1,J_2\}$} 
  & \multirow{2}{*}{ INS \cite{Mook1981}  }  
  & \multirow{2}{*}{87} 
  & \multirow{2}{*}{70} 
  & \multirow{2}{*}{69 \cite{McGuire1964}}  \\ 
FM, isotropic, 3D &  &  & & \\
 \hline\hline  
EuS, $S=7/2$ 
  & \multirow{2}{*}{$\{J_1,J_2\}$} 
  & \multirow{2}{*}{ INS \cite{Bohn1980}}  
  & \multirow{2}{*}{22} 
  & \multirow{2}{*}{15.3} 
  & \multirow{2}{*}{16.2 \cite{Moruzzi1963}}  \\ 
FM, isotropic, 3D &  &  & & \\
 \hline\hline  
MnF$_2$, $S=5/2$ 
  & \multirow{2}{*}{$\{J_1,J_2,J_3,D\}$} 
  & \multirow{2}{*}{ INS \cite{Morano2024} + ESR \cite{Hagiwara1996} }  
  & \multirow{2}{*}{87} 
  & \multirow{2}{*}{69} 
  & \multirow{2}{*}{67.7 \cite{Strempfer2004}}  \\ 
 AF, easy-axis, 3D &  &  & & \\
\hline\hline  
 MnTe, $S=5/2$
 & \multirow{2}{*}{$\{J_1,J_2,J_3,D\}$}
 &  INS \cite{Liu2024} 
 &  486 
 &  359 
 & \multirow{2}{*}{310 \cite{Komatsubara1963}} \\ 
\cline{3-5}  
 AF, easy-plane, quasi-1D  &
 & DFT \cite{Mazin2023} + ESR \cite{Dzian2025} 
 & 415 
 & 300  \\
\hline\hline  
 MnPSe$_3$, $S=5/2$, AF
 & $\{J_1,J_2,J_3,J_c\}$ 
 & powder INS \cite{Calder2021} 
 & 145 
 & 73 
 & \multirow{2}{*}{74 \cite{LeFlem1982}}  \\
\cline{2-5}  
 easy-plane XXZ, quasi-2D 
 & $\{J_1,J_1^{zz},J_2,J_3,J_c\}$ 
 & INS \cite{Liao2024} 
 & 119 
 & 77 & 
 \rule{0pt}{2.6ex} \\
\hline\hline  
    Rb$_2$MnF$_4$, $S=5/2$ 
  & \multirow{2}{*}{$\{ J_1, J_1^{zz} \}$} 
 & \multirow{2}{*} {INS \cite{Cowley1977}} 
 & \multirow{2}{*}{91} 
 & \multirow{2}{*}{41} 
 & \multirow{2}{*}{38.4 \cite{Birgeneau1970}}  \\
 AF, easy-axis XXZ, 2D & & & &  & \\
\hline\hline  
\multirow{2}{*}{\parbox[c]{3cm}{FePS$_3$, $S=2$ \\ AF, easy-axis, 2D}} 
 & $\{J_1,J_2,J_3,D\}$ 
 & INS~\cite{Lancon2016} 
 & 205
 & 142 
 &  \multirow{2}{*}{$118$ \cite{Ferloni1989}} \\ 
\cline{2-5}  
 & $\{J_{1a},J_{1b},J_2,J_3,D\}$ 
 &  INS \cite{Wildes2020} 
 & 212 
 & 130 
 & \rule{0pt}{2.6ex} \\
\hline\hline  
\multirow{2}{*}{\parbox[c]{3cm}{FePSe$_3$, $S=2$ \\ AF, easy-axis, 2D}} 
 &  $\{J_1,J_2,J_3,D\}$ 
 & ESR \cite{LeMardele2024} 
 & 151 
 & 108 
 & \multirow{2}{*}{$106$ \cite{Ferloni1989}} \\ 
\cline{2-5}  
 & $\{J_1,J_2,J_3,D\}$ 
 & INS \cite{Chen2024}  
 & 183 
 & 134 
 & \rule{0pt}{2.6ex} \\
\hline\hline  
  CoPS$_3$, $S=3/2$
 &  \multirow{2}{*}{$\{J_1,J_2,J_3,D,E\}$} 
 &  \multirow{2}{*}{INS \cite{Wildes2023}} 
 &  \multirow{2}{*}{153} 
 & \multirow{2}{*}{113} 
 &  \multirow{2}{*}{119 \cite{Wildes2017} } \\ 
AF, bi-axial, 2D &  &  &  &  \\
\hline\hline  
 CrSBr, $S=3/2$ 
 &  \multirow{2}{*}{$\{J_1,J_2,J_3,J_c,D,E\}$} 
 &  \multirow{2}{*}{INS \cite{Scheie2022} + ESR \cite{Cho2023}} 
 &  \multirow{2}{*}{300} &  \multirow{2}{*}{$166\pm 3$} 
 &  \multirow{2}{*}{132 \cite{Goser1990} } \\
 AF, bi-axial, quasi-2D  &  &  &  & \\
\hline\hline  
 CrI$_3$, $S=3/2$ 
 &  \multirow{2}{*}{$\{J_1,J_2,J_3,J_{c1},J_{c2},D\}$} 
 &  \multirow{2}{*}{INS \cite{Chen2021}} 
 &  \multirow{2}{*}{97}  
 &  67 (bulk) 
 &  61  \cite{McGuire2015,Lin2018} \\ 
 FM, easy-axis, quasi-2D &  &  &  & 46 (ML)  &  45 \cite{Huang2017} \\
\hline\hline   
 \end{tabular}
\caption{Computed  and measured transition temperatures for selected magnetic materials.
The first column presents the chemical formula, spin value, type of anisotropy and dimensionality.
The second and third columns provide the microscopic spin model and method used to obtain 
the interaction parameters. Here, INS and ESR denote the inelastic-neutron scattering  and 
the electron-spin resonance experiments, whereas DFT stands for the density-functional calculations.  
The last three columns list transition temperatures from mean-field calculations $T^\text{MF}_c$, 
classical Monte Carlo simulations $T^\text{MC}_c$ (both this work) and experimentally measured 
$T^\text{exp}_c$.  Unless specified, the accuracy of $T_c$ value is less than the last shown digit.
}
\label{TMC}
\end{table*}
\end{center}
 
In contrast to the case of MnF$_2$,  the cumulants for FePSe$_3$ exhibit a non-monotonic  dependence on 
temperature with a peak singularity for $L\to\infty$. The nonmonotonic behavior of the fourth-order cumulants 
signifies the first-order transition \cite{Challa1986,Vollmayr1993,Jin2012}. Such a dependence of $U_4(T,L)$  
is related to a double peak structure in the  probability distribution of the order parameter characteristic for 
the phase coexistence \cite{Challa1986}. By extrapolating the peak position versus $1/N\to 0$, we determine 
the transition temperature $T_c = 107.8\pm 0.1$~K.  The natural question is:  why does the sixfold ($2n=6$) 
symmetry breaking in FePSe$_3$ occur via a first-order transition?

Generally, phase transitions with the $Z_n$ symmetry breaking can belong to either the $n$-state Potts or  
the $n$-state clock universality classes. A simple geometrical analysis of the domain walls between the $C_3$ 
rotated zigzag antiferromagnetic states suggests a similarity with the Potts model. Furthermore, the six-state 
Potts model has a first-order transition in 2D \cite{Wu1982}, whereas the six-state clock model exhibits two 
successive Kosterlitz-Thouless  transitions \cite{Jose1977}. Hence, we conclude that FePSe$_3$  provides 
a rare realization of the 2D spin system in the  six-state Potts universality class \cite{Domany1978}. The recent 
experimental study has clearly demonstrated the first-order nature of the phase transition in  FePSe$_3$ 
\cite{Chen2024} in full agreement with our theoretical result.
 
Monte Carlo simulations and their detailed data analysis  have been performed for each of the selected 
materials. The computed transition temperatures are presented in Table~\ref{TMC}.  In addition, Table 
provides a brief information about the spin model for each material  with corresponding references, see 
Appendix~\ref{Models} for further details. Together with the MC result $T_c^{\rm MC}$ and the measured  
value $T_c^{\rm exp}$ we also include a mean-field  estimate for the transition temperature:
\begin{equation}
k_BT_c^{\rm MF} = -\frac{1}{3} S(S+1) \sum_{\bm{r}} J_{ij} e^{-i\bm{Q}\cdot\bm{r}_{ij}} \ .
\label{MF}
\end{equation}
where $\bm{Q}$ is the ordering wavevector and $\bm{r}_{ij}$ are intersite distances, see, for example, 
\cite{Johnston2011,MZH2022}. As expected, the mean-field approximation significantly overestimates 
the transition temperatures. The actual error varies widely, from 25\%\ for the 3D spin model of MnF$_2$ 
with a high coordination number  of $z=8$, to 100\%\ in the case of the 2D model of Rb$_2$MnF$_4$ with 
a low  $z=4$ and weak easy-axis anisotropy.

Now, let us compare Monte Carlo results for transition temperatures with their experimental values. 
For majority of studied models, the agreement between theory and experiment is quite good with difference 
between the two values of the order of 3--6\%. 
We reiterate that, for majority of quoted examples, the microscopic input parameters were derived  from the low-temperature 
spectroscopic data fitted to magnon energies for spin-$S$ models. 
Meanwhile, our simulations were performed using classical vector models with $S_C=\sqrt{S(S+1)}$,
which already makes the observed agreement nontrivial.
The most spectacular match is found 
for the spin-7/2 ferromagnet EuO and the spin-5/2 collinear antiferromagnet MnF$_2$. For both of them
the discrepancy of calculated and measured $T_c$'s is less than  2\%.
The computed and measured transition temperatures are 
also in a good agreement for FePSe$_3$ for a set of interaction constants from \cite{LeMardele2024}.
The correspondence worsens for an alternative set  from the INS study \cite{Chen2024}. 
Of course, we cannot claim that one set of interaction constants is superior to another based solely on
 $T_c$ values. However, calculated transition temperatures provide an additional criterion for selecting 
 among multiple sets of microscopic parameters.

A good agreement between theory and experiment is also found for EuS, MnPSe$_3$, Rb$_2$MnF$_4$, 
CoPS$_3$, and the monolayer of CrI$_3$, see Table~\ref{TMC}.  The largest difference  between two $T_c$'s of the order of 25\%\
is obtained for CrSBr.  Regarding a possible role of quantum effects 
in this mismatch, we note that the computed $T_c^{\rm MC}=166$~K exceeds the strongest in-plane exchange $J_2S(S+1)=147$~K
 \cite{Scheie2022}. In this temperature range, the quantum corrections for a similar 2D spin-3/2 model 
 do not exceed 3--5\% both for the specific heat and the susceptibility \cite{Johnston2011}. Hence, 
 uncertainty of interaction constants for CrSBr is a more plausible source of this disagreement.

However, quantum corrections can be significant near the transition temperature in the 2D easy-axis 
antiferromagnet Rb$_2$MnF$_4$. As  discussed in more detail in Appendix~\ref{Rb2MnF4}, the computed ordering temperature
for Rb$_2$MnF$_4$ is $T_c^{\rm MC} \simeq 0.6 JS(S+1)$. At this temperature, the difference between  
quantum and classical MC simulations results for the spin-5/2 square-lattice antiferromagnet are at the level of 5\%  for 
the magnetic susceptibility and increase to 10--15\% for the specific heat  \cite{Johnston2011}. Therefore, quantum 
effects should be significant and Rb$_2$MnF$_4$ can serve as an experimental test bed for future studies of these
effects in 2D magnets.

\begin{figure}[tb]
\centerline{
\includegraphics[width=0.9\columnwidth]{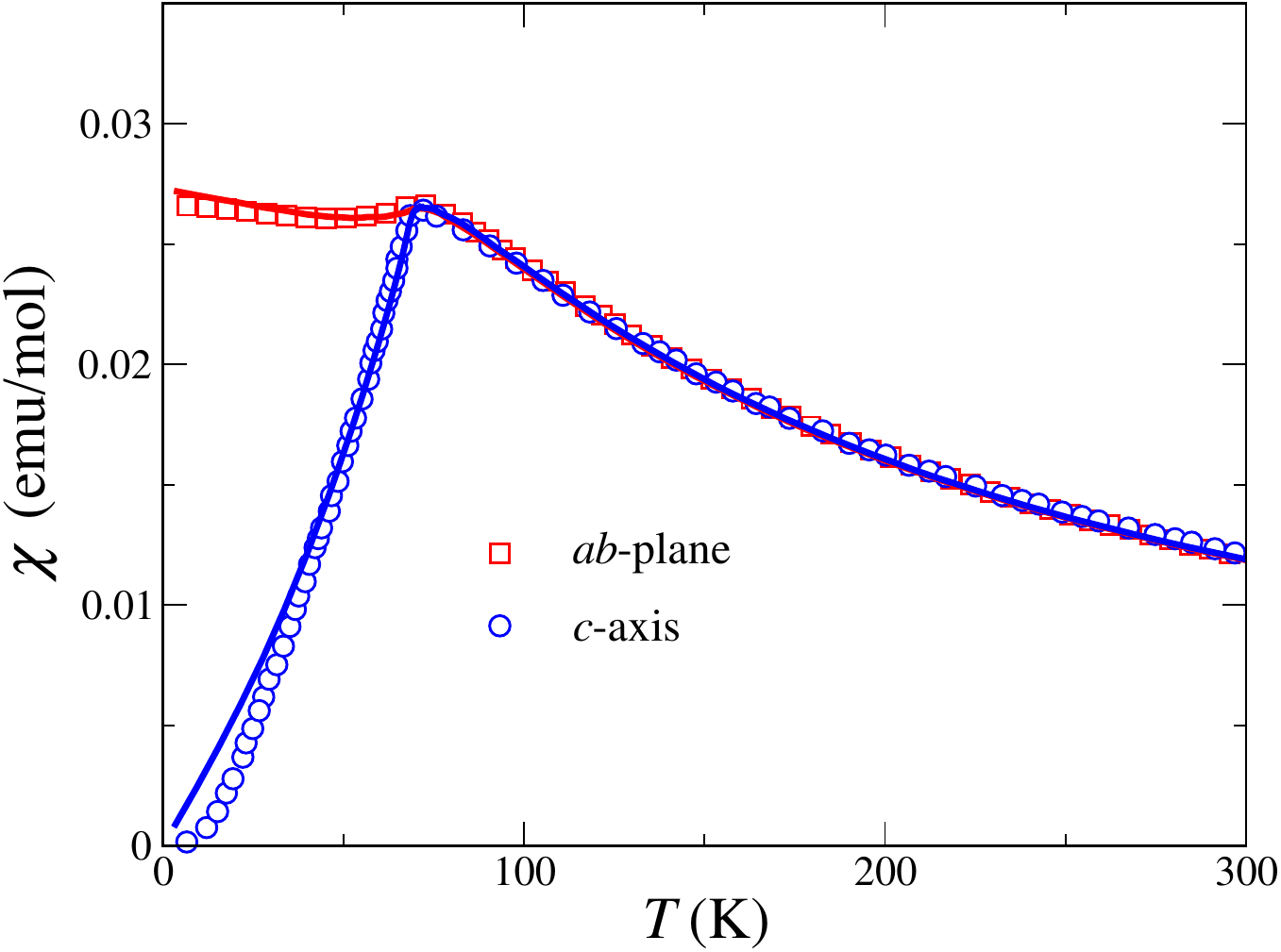}
}
\caption{
Magnetic susceptibility of MnF$_2$ along two principal directions.
The symbols represent experimental data from Ref.~\cite{Schleck2010} and the lines are theoretical Monte Carlo results for the 
spin model detailed in Appendix \ref{MnF2}.
}
 \label{Chi}
\end{figure}

Classical MC simulations can be used not only to obtain transition temperatures for realistic spin models, 
but also to compute other  physical properties of magnetic materials.  In  Figure~\ref{Chi},
theoretical results for the magnetic susceptibility of the easy-axis collinear antiferromagnet 
MnF$_2$ are compared with experimental data  
\cite{Schleck2010}.  
Good agreement is found both in the disordered paramagnetic phase, $T>T_c$, 
and in the antiferromagnetic state for $0.5T_c \alt T < T_c$. In the ordered phase, the susceptibility of a collinear 
antiferromagnet exhibits a characteristic splitting between the longitudinal $\chi_\parallel$ and transverse $\chi_\perp$
components. The sublattice orientation in MnF$_2$ is determined by a weak easy-axis anisotropy along 
the $c$ axis, which has an almost negligible effect in the paramagnetic state at $T>T_c$.
 The expected susceptibility behavior emerges correctly in our MC simulations. However, the low-temperature 
asymptotes determined by quantum statistics cannot be reproduced within a  classical description
and the corresponding discrepancy with experimental data for MnF$_2$ becomes apparent below $\sim 30$~K.
Nevertheless, the measured  values  of $\chi(T)$ around and below $T_c$ are remarkably well reproduced by 
the MC simulations. Similar quantitative agreement can be expected for other thermodynamic properties, such as the heat 
capacity $C(T)$ and the magnetocaloric response $(\partial S/\partial H)_T= (\partial M/\partial T)_H$.

\section{Conclusions}
\label{Sum}

We establish the mapping  between the statistical mechanics of  a quantum spin system 
and  an effective  classical model. We show that in the large-$S$ limit, the asymptotic form of the quantum 
partition function $Z_Q(S,T)$ coincides with the classical one $Z_C(S_C,T)$ for spins of length $S_C=\sqrt{S(S+1)}$. 
The obtained mapping sets a framework  for a systematic expansion in powers of $1/S_C^2$ for the thermodynamic properties 
of a quantum spin model. 
The first, leading term of any physical observable is given by a spherical integral and can therefore be evaluated
using classical MC simulations. 
We obtain classical expressions for the standard terms in the spin Hamiltonian (\ref{map_MC}). 
In particular, the mapping of the magnetic anisotropy constant contains a universal quantum correction. 
An important consequence of the presented theory 
is the leading $S(S+1)$ scaling of the transition temperature for general spin models consistent with the phenomenological
form (\ref{Tc}) suggested by numerical studies of the nearest-neighbor Heisenberg model \cite{Oitmaa2004,Oitmaa2018}.

To demonstrate the predictive power of this approach, we performed classical MC simulations 
for several magnetic materials, with obtained transition temperatures summarized in Table~\ref{TMC}.
An interesting direction for future studies would be to extend the mapping scheme by computing quantum
 $1/S(S+1)$  corrections to the classical behavior for $T\geq T_c$.

Finally, let us mention that other types of quantum-to-classical crossover are also possible in quantum magnetism.
One example is anisotropic spin models with dominant uniaxial interactions. In this case the crossover to the classical
Ising behavior is expected to take place even for small $S=1/2$ for weak enough strength of transverse coupling.
Metallic materials with  local magnetic moments partially screened by conduction electrons may 
provide another example. In such systems, the classical behavior can be rooted in 
the mean-field like character of the many-body magnetic state.

\acknowledgements

We are grateful  to  A. Honecker, D. V. Kveshchenko, M. Orlita,  and B. Parisse 
for valuable discussions.  We  also thank 
L. Messio,
O. A. Petrenko, and T. Ziman for comments and suggestions 
on the manuscript.

\appendix
\section{Asymptotic form of spin traces}
\label{Trace}

In this Appendix we show how the asymptotic form of spin traces
 $I_n(S)$ can be obtained without Bernoulli polynomials using only the elementary mathematical functions. 
Introducing the generating function  
\begin{equation}
 I(S,t) = \frac{1}{2S+1}\sum_{m=-S}^{S} e^{mt} \ ,
 \label{ISt}
\end{equation} 
we can relate  the spin traces  to  its derivatives at $t=0$:
\begin{equation}
I_n(S) = \left.\frac{\partial^n}{\partial t^n}I(S,t)\right|_{t=0} \,.
\end{equation} 
The geometric series summation yields
\begin{equation}
 I(S,t)  =   \frac{1}{2S+1} \frac{\sinh(S+\frac{1}{2})t}{\sinh \,t/2} = f(t)\,g(t)  \,, 
 \label{ISfin}
\end{equation}
where 
\begin{equation}
  f(t) =  \frac{\sinh(S+\frac{1}{2})t}{(S+\frac{1}{2})t} =   \sum_{p=0}^\infty \frac{[(S+\frac{1}{2})t]^{2p}}{(2p+1)!}
\end{equation}
and $g(t) =  (t/2)/\sinh(t/2)$. Both $f(t)$ and $g(t)$ are even functions  and their odd derivatives vanish 
at $t=0$. Moreover,
\begin{equation}
  f^{(2k)}_t(0) =  \frac{(S+\frac{1}{2})^{2k}}{2k+1}\,.
\label{f2k0}  
\end{equation}

The binomial expansion of the $n$-th derivative of  (\ref{ISfin}) yields $I_n(S)$ as a polynomial of 
degree $n/2$ in the variable $Z=(S+1/2)^2$. Then, a substitution $Z = X + 1/4$ is  used to rewrite it 
as a polynomial in $X=S(S+1)$.  Since $g(0)=1$, the highest degree term is  directly given by (\ref{f2k0}).
Altogether, we obtain the polynomial representation
\begin{equation}
I_n(S) = \frac{X^{n/2}}{n+1}  + \sum_{k=1}^{n/2-1} a_k X^k  \ .
\end{equation}
For $S=0$, the generating function $I(S,t)$ is a constant. Hence, its derivatives expressed as polynomials
of $X$ must vanish for $X=0$. This explains the absence of a constant term in the above polynomial.

\section{Generalized Marsaglia algorithm}
\label{Marsaglia}

Standard techniques for sampling points from a two-dimensional sphere either require computing trigonometric 
functions or use three random numbers per point  \cite{Landau2000}. Marsaglia has suggested an elegant method 
that generates a uniform distribution of points on a sphere using only two random numbers and no transcendental functions  \cite{Marsaglia1972}. However, in its original formulation this algorithm is not well-suited for 
MC simulations of spin models, which generally require restricted spin moves over a sphere.  Here, we describe 
a modified version of Marsaglia algorithm that generates points on a spherical cap of height $h$.

Suppose the random numbers are uniformly distributed  in the interval $x \in (-1,1)$. The rejection method can be 
used to generate random points $(x_1,x_2)$ uniformly distributed on a disk by requiring  $x_1^2 + x_2^2 \leq 1$.
The two derived parameters $r = x_1^2 + x_2^2$ and  $\phi =\arg (x_1+ix_2)$  are uniformly distributed as well: 
$r \in (0,1)$ and $\phi \in (0,2\pi)$. Assigning $\phi$ to the azimuthal angle of a spin, we express its $z$-component as  
\begin{equation}
S^z  =  1- hr \ ,
\end{equation}
which has a uniform  distribution on  $(1-h,1)$.
Then, the transverse spin components are given by
\begin{eqnarray}
S^x  & = & \cos(\phi)\sqrt{1-S^{z2}} =  x_1\sqrt{h(2-hr)}\,, \nonumber \\
S^y  & = & \sin(\phi)\sqrt{1-S^{z2}} =  x_2\sqrt{h(2-hr)}
\label{Sxy}
\end{eqnarray}
and, hence, can be computed using only a square-root function. For $h=2$, the above equation coincides with 
the original formula given by Marsaglia \cite{Marsaglia1972}.  The restricted spin moves used in MC 
simulations can be generated by fixing the height of a polar cap to  the simulation temperature: $h\simeq T$.

\section{Spin models for selected materials}
\label{Models}
 
\subsection{EuO(S)}
\label{EuO}

Europium chalcogenides EuX  (X = O, S, Se, Te)  have been investigated since the early 1960s due to 
remarkable optical, magnetic, and transport properties \cite{Mauger1986}.  
Among them, EuO and EuS are old examples of ferromagnetic insulators,
 with Curie temperatures of  69~K \cite{McGuire1964} and
16.2~K \cite{Moruzzi1963}, respectively.  The two materials share the rock-salt crystal structure, where Eu$^{2+}$
 ions reside on the sites of a face-centered cubic lattice (fcc). The $4f$ europium moments have vanishing 
 orbital moment $L=0$ and large spin $S=7/2$, which leads to predominantly isotropic magnetic interactions. 
 The principal interactions in these materials are two nearest-neighbor exchanges. The inelastic neutron-scattering (INS) measurements on single crystals yield $J_1 = -1.25$~K and 
$J_2 = -0.25$~K for EuO \cite{Mook1981} and $J_1 = -0.442$~K and $J_2 = 0.2$~K for EuS \cite{Bohn1980}. 
Note that according to the chosen parameterization (\ref{Hgen}), a negative sign corresponds to ferromagnetic exchange.

We have performed the classical Monte Carlo simulations of the $J_1$--$J_2$ Heisenberg model on the fcc lattice
using the above values of the exchange constants and obtained the transition temperatures 
$T_c^{\rm MC}= 70$~K for EuO and  $T_c^{\rm MC} = 15.3$~K for EuS in good correspondence with the
measured transition temperatures. To further improve agreement between theory and experiment, one may include 
 the dipole-dipole interactions, which amount to 4\%\ of the nearest-neighbur exchange $|J_1|S^2$ for EuO
and 7.5\%\ for EuS.

\subsection{MnF$_{\bf 2}$}
\label{MnF2}

MnF$_2$ is one of the best known examples of the  collinear Heisenberg  antiferromagnet with large spins 
$S=5/2$ of Mn$^{2+}$ ions. It has the body-centered tetragonal crystal structure. Spins in the corners and in 
the centre of the tetragonal unit cell order in the opposite directions along the $c$ axis below the  transition 
temperature $T_c\simeq 67.7$~K. The microscopic interaction parameters of MnF$_2$ were recently refined 
by Morano {\it et al.}~\cite{Morano2024}  with  the following values for three dominant exchanges:  
$J_1 = -0.0677$~meV, $J_2 = 0.3022$~meV, and $J_3 = -0.0044$~meV. In addition, 
the magnetic anisotropy constant was estimated as $D = -0.0267$~meV. The easy-axis  anisotropy is responsible for 
the orientation of the magnetic sublattices along the $c$ axis as well as appearance of a finite magnon gap:
\begin{equation}
\Delta = 2S\sqrt{|D|(J_{\rm eff}+|D|)}\ ,
\label{DEA}
\end{equation}
where $J_{\rm eff} =8J_2$ is a net exchange  coupling between  the two opposite sublattices. 
For the above set of microscopic parameters, one finds   $\Delta = 1.277$~meV or 308.8~GHz,
which is somewhat larger than  the  ESR gap $\Delta=259.7$~GHz  \cite{Hagiwara1996}. Since, 
the ESR technique is superior over the inelastic neutron scattering  at low energies, we used the ESR gap 
value $\Delta = 1.074$~meV to refine the anisotropy constant to $D = -0.0189$~meV. 

The above value of $D$ together with the quoted  exchange parameters from the inelastic neutron-scattering
experiments have been used in our Monte Carlo simulations.  
We find a good agreement between  the experimental $T_c^{\rm exp}\approx 68$~K and  the
computed $T_c^{\rm MC}=69$~K transition temperatures.

\subsection{MnTe}
\label{MnTe}

MnTe is an easy-plane collinear antiferromagnet, which recently attracted attention due to the experimental 
observation of `altermagnetic' splitting of the magnon bands \cite{Liu2024}.  
It has the hexagonal crystal structure $P6_3/mmc$ and orders  at $T_c \simeq  310$~K. 
In the antiferromagnetic state, manganese spins, $S=5/2$ are parallel  inside triangular layers and have the opposite orientation 
between adjacent layers. According to the INS measurements  \cite{Liu2024}, the principal exchanges are 
the first-neighbor coupling between layers $J_1=3.99$~meV, the nearest-neighbor exchange within each layer $J_2=-0.12$~meV, 
and the next nearest-neighbor exchange between layers $J_3=0.472$~meV. In addition, the single-ion anisotropy of the easy-plane 
type has been determined as $D=0.0482$~meV. An alternative set of microscopic parameters
has been obtained from the density functional (DFT) calculations: 
$J_1=3.628$~meV, $J_2=0.078$~meV, and  $J_3=0.457$~meV  \cite{Mazin2023}.  
We have used both  sets of the exchange constants in our MC simulations. For the DFT  set of exchanges, we add
the single-ion anisotropy constant derived from the ESR magnon gap $\Delta \approx 3.5$~meV \cite{Dzian2025} 
using
\begin{equation}
\Delta =  2S\sqrt{DJ_{\rm eff}} \ ,
\end{equation}
where $J_{\rm eff} =2J_1 + 12J_3$ is a net exchange  coupling between the opposite magnetic sublattices.
The above equation yields $D=0.0385$~meV, which is smaller  than the INS estimate.

The transition temperature derived from the MC simulations for the DFT set of parameters
$T_c^{\rm MC} = 300$~K  gives a better match of the measured $T_c=310$~K 
in comparison  to $T_c^{\rm MC} \simeq 360$~K computed with the interaction constants from the INS measurments.
Due to the presence of at least four microscopic parameters in the spin model of MnTe, we cannot 
 claim superiority of one set over the other. Instead, the example of MnTe supports 
our general conclusion that MC simulations  of transition temperatures for magnetic materials can be used  to scrutinize any suggested
set of microscopic parameters obtained from experiments or  ab initio calculations.

\subsection{R\lowercase{b}$_{\bf 2}$M\lowercase{n}F$_{\bf 4}$}
\label{Rb2MnF4}

The square-lattice antiferromagnet Rb$_2$MnF$_4$ orders at $T_c=38.4$~K into a simple N\'eel structure \cite{Birgeneau1970}.  
The crystal lattice  structure  frustrates exchange coupling between Mn layers and it is reasonable to consider
this material as a pure 2D magnet. A weak magnetic anisotropy of Rb$_2$MnF$_4$  is often approximated using 
the nearest-neighbor anisotropic-exchange model:
\begin{equation}
 \hat{\mathcal{H}} =\sum_{\langle ij\rangle} \Bigl[  J\bigl(S^x_i S^x_j+S^y_iS^y_j\bigr) + J^{zz} S^z_iS^z_j \Bigr] \,.
\label{HXXZ}    
\end{equation}  
Specifically, Cowley {\it et al.}~\cite{Cowley1977} estimated
$J=0.654$~meV and $J^{zz}/J = 1.005$ from fits to the INS spectra. Using these parameters, we obtained
the transition temperature 
 $T_c=41$~K, which is a fraction of the `classical' exchange:  $T_c\simeq 0.6 JS(S+1)$.
 Based on the comparison of the QMC and CMC results for the Heisenberg spin-5/2 square-lattice antiferromagnet 
\cite{Johnston2011}, a quantum correction of 5--10\%\ to $T_c$ can be expected in this case,
which may explain the difference  with the measured value  $T_c = 38.4$~K.
Besides, for a better description 
of the experimental properties of Rb$_2$MnF$_4$ one may include the long-range dipolar interactions instead of an effective  
anisotropic exchange assumed by  Eq.~(\ref{HXXZ}).

.
\subsection{F\lowercase{e}PS$_{3}$}
\label{FePS3}

Antiferromagnet FePS$_3$ belongs to a large family of layered MPX$_3$  compounds with ${\rm X} = {\rm S}$, Se 
 \cite{LeFlem1982,Brec1986}. 
Recently, there was a surge of interest in FePS$_3$ due to the discovery
of antiferromagnetic ordering in monolayer samples of this material \cite{Lee2016}. The presence of a finite-temperature
transition in a pure 2D spin system becomes possible due to a very large easy-axis anisotropy induced by the crystalline electric field 
for the $S=2$ iron magnetic moments.  Consequently, an interlayer coupling is often neglected
considering FePS$_3$ as a 2D antiferromagnet on an ideal honeycomb lattice.

Lan{\c c}on {\it et al.}~\cite{Lancon2016}  used linear spin-wave
fits of the measured magnon dispersion in FePS$_3$. They deduced
in-plane exchange constants up to the third neighbors: $J_1 =  -2.92$~meV,   $J_2 =  0.08$~meV,  $J_3 =  1.92$~meV
together with the single-ion anisotropy $D=-2.66$~meV.
The interplay between two competing exchanges  $J_1$ and $J_3$ stabilizes
 the zig-zag antiferromagnetic structure of iron spins. 
For collinear easy-axis antiferromagnets at $T=0$,
 the single-ion anisotropy constant $\bar{D}$ that enters the expression for magnon energies  is renormalized 
 from the bare value $D$  to
 $\bar{D} = D[1-1/(2S)]$, see \cite{Oitmaa2008,ElMendili2025}.  
 Monte Carlo simulations are performed for the microscopic model with the bare $D$.
Accordingly, we have used  $D=4D_{\rm exp}/3 =-3.55$~meV in Eq.~(\ref{map_MC}) for the effective
classical  model of FePS$_3$.

Improved  theoretical fits of the same magnon dispersion data for FePS$_3$ have been obtained 
by Wildes {\it et al.}~\cite{Wildes2020}
 by allowing for inequivalent nearest-neighbor couplings
 $J_{1a}$ and $J_{1b}$ for parallel and antiparallel pairs of iron spins.
 The microscopic parameters from the new fits are
$J_{1a} =  -2.9$~meV,  $J_{1b} = -0.7$~meV, $J_2 =  0.094$~meV,  $J_3 =  1.28$~meV, and  $D=-2.53$~meV.
The suggested explanation for such significant difference between $J_{1a}$ and $J_{1b}$ is the presence of a 
biquadratic exchange 
\begin{equation}
 \hat{\mathcal{H}}_{\rm biq} = -K_1 \sum_{\langle ij\rangle} \bigl(\bm{S}_i\cdot \bm{S}_j\bigr)^2  \,,
\label{Hbiq}    
\end{equation} 
which in the collinear antiferromagnetic states at $T=0$ produces $J_{1a,b} = J_1 \mp 2K_1S^2$
for parallel/antiparallel spins.
 We have simulated the phase transition in FePS$_3$ using both
 $J_{1a}$--$J_{1b}$ and  $J_1$--$K_1$ spin models. 
Table~\ref{TMC}   includes the result of  the former model, which 
agrees  better with the experimental $T_c$.

\subsection{F\lowercase{e}PSe$_{3}$}
\label{FePSe3}

FePSe$_3$ is a sister material of  FePS$_3$. The transition temperatures and microscopic values of the in-plane
exchange constants are similar for both materials. The infrared absorption measurements by Le Mardel\'e {\it et al.}~\cite{LeMardele2024} have given the following set of microscopic constants for FePSe$_3$: $J_1 = -2.5$~meV, $J_2 = 0.5$~meV, $J_3 = 1.0$~meV and $D = -3.7$~meV. An alternative  set of the parameters has been obtained in the INS studies \cite{Chen2024}: $J_1 = -2.3$~meV, 
$J_2 = 0.23$~meV, $J_3 = 2.01$~meV and $D = -2.74$~meV.  Again, ferromagnetic $J_1$ and  antiferromagnetic $J_3$ 
are the strongest in-plane exchanges. However, in contrast to FePS$_3$, the second-neighbor exchange  $J_2$
significantly increases in the selenium compound.  Our MC simulations of the 2D model for FePSe$_3$ with the
 two sets of microscopic constants yield, respectively, transition temperature at $T_c=108$~K and 134~K 
 that have to be compared to the  measured value  of $T_c =106$~K \cite{Ferloni1989},  see Table~\ref{TMC}.
Using the spin model of  FePSe$_3$, we have also verified  the role of quantum renormalization for the magnetic anisotropy
constant  (\ref{DC}). For $S=2$,  the quantum correction reduces the classical single-ion  constant  by $\sim 13$\%.
Performing MC simulations for the microscopic set \cite{LeMardele2024}  without such a correction
 we have obtained $T_c\approx 113$~K. This result is notably  
further  from the experimental value, $T_c^{\rm exp} =106$~K,  than the quoted  $T_c^{\rm MC} = 108$~K computed
with the correctly rescaled $D$.

\subsection{M\lowercase{n}PS\lowercase{e}$_{\bf 3}$}
\label{MnPS3}

The hexagonal  layers of manganese spin-5/2 ions in MnPSe$_3$  are stacked in a period-3 structure
along the $c$ axis. The collinear antiferromagnetic state in MnPSe$_3$ has a simple N\'eel structure indicating a dominant 
nearest-neighbor exchange $J_1>0$.  In our MC simulations we have employed two sets of microscopic interaction
 parameters for MnPSe$_3$. The first set was suggested by Calder {\it et al}.~\cite{Calder2021} based on the powder INS
data. It includes the in-plane exchange constants 
up to the third neighbors: $J_1 =  0.9$~meV,   $J_2 =  0.06$~meV,  $J_3 =  0.38$~meV, 
as well as a weak interlayer coupling $J_c = 0.062$~meV. The second set of parameters was put forward by 
Liao {\it et al}.~\cite{Liao2024}, who performed INS experiments with single crystals of MnPSe$_3$:
 $J_1 =  0.73$~meV,   $J_2 =  0.017$~meV,  $J_3 =  0.43$~meV, and $J_c = 0.054$~meV.
In addition, these authors have included  a weak $XXZ$ anisotropy  on the strongest nearest-neighbor
bonds with $J^{zz}_1/J_1=0.981$. The two sets of microscopic  constants yield  transition temperatures of $T_c=73$~K and 77~K, 
which are both close to the experimental value $T_c =74$~K \cite{LeFlem1982}.

\subsection{C\lowercase{o}PS$_{3}$}
\label{CoPS3}

Wildes {\it et al.}~\cite{Wildes2023} measured spin waves in the single crystals of  CoPS$_3$ in the INS
experiments.
They fitted their data using the microscopic spin-3/2 model with three in-plane exchange constants
and the bi-axial magnetic anisotropy
\begin{equation}
\label{CoPS3}
 \hat{\mathcal{H}}_{\rm SI}
  = \sum_i  \Bigl[ D{S_i^z}^2 + E  \bigl( {S_i^x}^2\! - {S_i^y}^2 \bigr) \Bigr] \,.
\end{equation}
Presence of the bi-axial term is in agreement with the low monoclinic symmetry of CoPS$_3$ crystals. 
Linear spin-wave theory yields $J_1 = -1.37$~meV, $J_2 = 0.09$~meV,  and $J_3 = 3.0$~meV with
$D = 6.07$~meV and $E = -0.77$~meV for the magnetic anisotropy. The large, positive value of $D$ indicates
a strong easy-plane anisotropy in this material. In this case, there is no universal renormalization of  bare $D$ by a factor of $(1-1/2S)$ 
in the expression for a magnon gap at $T=0$ \cite{Oitmaa2008}. On the other hand,
 in the paramagnetic phase at $T>T_c$, the leading quantum correction to anisotropy is given by Eq.~(\ref{map_MC}),  
 which must be applied to both  $D$ and $E$ constants.
The transition temperature obtained in
MC simulations with the aforementioned set of microscopic constants is $T_c= 113$~K, which is in good agreement with the measured 
$T_c= 119$~K \cite{Wildes2017}.

\subsection{C\lowercase{r}SB\lowercase{r}}

CrSBr is a spin-3/2 orthorhombic van der Waals material. It magnetically orders at $T_c = 132$~K \cite{Goser1990}  with a parallel 
arrangement of chromium moments within the $ab$ layers, which are stacked antiferromagnetically in the $c$ direction. 
 INS measurements \cite{Scheie2022}  have determined the in-plane exchange interactions 
up to eight neighbors with the
three principal constants $J_1 = -1.9$~meV, $J_2 = -3.38$~meV, and  $J_3 = -1.67$~meV.
The other in-plane exchange parameters are small, not exceeding 5-10\%\ of the largest $J_2$ coupling. 
However, the accuracy of the neutron experiments is insufficient to determine much weaker interlayer exchange and anisotropy constants. 
These parameters were measured in ESR experiments   \cite{Cho2023}.
The interlayer exchange coupling is estimated to be $J_c = 0.0694$~K or 6.0~$\mu$eV and
the bi-axial magnetic anisotropy parameters are $D_b=-0.3956$~K and $E_{ca}=0.2074$~K. 
Here, $D_b$ describes the easy-axis anisotropy in the $b$ direction, while $E_{ca}$ parameterizes anisotropy between the hard $c$ direction and the intermediate $a$ axis. The hierarchy of anisotropy scales becomes clearer if one uses instead the parameterization
similar to ({\ref{CoPS3})  with $D_c$ and $E_{ab}$. A simple geometric transformation yields
$D_c = 0.509$~K  or   $43.9$~$\mu$eV  and
$E_{ab} = 0.094$~K  or $8.1$~$\mu$eV. Thus, the magnetic anisotropy of CrSBr is  predominantly of  the easy-plane type with 
a small in-plane anisotropy between the $a$ and $b$ directions. 
The weak magnetic anisotropy and tiny inter-plane exchange  are 
essential for establishing a non-zero  transition  temperature in CrSBr. In MC simulations, the  transition takes place at
$T_c = 166\textrm{K}\simeq  J_2S(S+1)=147$~K. The quantum effects in the corresponding temperature range are  small a priori.
Therefore, the discrepancy between the experimental value, $T_c = 132$ K, and the simulation result, $T_c = 166$~K, may stem from 
uncertainties in the in-plane exchange constants.

\subsection{C\lowercase{r}I$_{3}$}

CrI$_3$ is a prominent example of van der Waals honeycomb-lattice material, which exhibits ferromagnetism down to the
monolayer limit \cite{Huang2017}. Treated as a promising candidate for spintronics applications, CrI$_3$ has also attracted attention 
due to a possible realization of topological magnon bands 
\cite{Kim2016}. We use in our MC simulations of CrI$_3$
the following set of microscopic  constants :
$J_1 = -2.11$~meV,  $J_2 = -0.11$~meV, and  $J_3 = 0.1$~meV for the intra-layer exchanges, the anisotropy
constant $D = -0.123$~meV and two inter-layer exchanges  $J_{c1} =  0.048$~meV and $J_{c2} = - 0.071$~meV  \cite{Chen2021}.
The single-ion anisotropy plays an essential role in stabilizing the Ising-type of ferromagnetic order in 2D layers. We rescaled
the experimental value of $D$ using the same two-step renormalization procedure as discussed in the subsection for FePS$_3$. 
For simulations of the phase transition in the mono-layer sample we used the same bulk values of the in-plane exchange constants. 
Ferromagnetic transition for a single layer is reproduced with very good accuracy. An error of 10\%\ between theoretical and experimental $T_c$ for bulk samples may be assigned to the remaining uncertainty in the values of inter-plane exchange.

\bibliography{QCC}

\end{document}